\documentclass[12pt]{article}

\usepackage{lipsum}

\usepackage{xr-hyper}
\usepackage{amsthm}
\usepackage{fullpage}
\usepackage{footnote}
\usepackage{tabu}
\usepackage{booktabs}
\usepackage[title]{appendix}
\usepackage{scribe}
\usepackage{caption}
\usepackage{mathtools, nccmath}

\providecommand{\keywords}[1]
{
  \small	
  \textbf{\textit{Keywords---}} #1
}
\usepackage{etoolbox}

\makeatletter
\patchcmd{\@makecaption}
  {\parbox}
  {\advance\@tempdima-\fontdimen2} % decrease the width!
  {}{}
\makeatother 

\usepackage{titling} 

\usepackage{amsmath,bbm,amssymb, amsfonts}
\usepackage{mathtools}
\usepackage{graphicx,psfrag,epsf}
\usepackage{natbib}
\usepackage{url} % not crucial - just used below for the URL 
\usepackage{color}
\usepackage{algorithmicx}
\usepackage{booktabs}
\usepackage{siunitx}
\usepackage{array}
\usepackage{rotating}
\usepackage{adjustbox}

\usepackage{algorithm}
\usepackage{algpseudocode}

\usepackage{tikz}

\usepackage{url}

\usepackage{multicol}
\usepackage{xcolor}
\usepackage{algorithm}
\makeatletter
\def\breve{\mathpalette\wide@breve}
\def\wide@breve#1#2{\sbox\z@{$#1#2$}%
	\mathop{\vbox{\m@th\ialign{##\crcr
				\kern0.08em\brevefill#1{0.8\wd\z@}\crcr\noalign{\nointerlineskip}%
				$\hss#1#2\hss$\crcr}}}\limits}
\def\brevefill#1#2{$\m@th\sbox\tw@{$#1($}%
	\hss\resizebox{#2}{\wd\tw@}{\rotatebox[origin=c]{90}{\upshape(}}\hss$}
\makeatletter

\newtheorem{theorem}{Theorem}

\def\gobblestop#1#2{#1}
\def\killstop{%
	\aftergroup\gobblestop
}
\def\thick#1{\hbox{\rlap{$#1$}\kern0.25pt\rlap{$#1$}\kern0.25pt$#1$}}

\def\smbalpha{\boldsymbol{{\scriptstyle{\alpha}}}}

\def\smbalpha{\widehat{\smbalpha}}

\def\hbar{\bar{ h}}

\def\mybox#1{\vskip1mm \begin{center}
        \hspace{.0\textwidth}\vbox{\hrule\hbox{\vrule\kern6pt
\parbox{.9\textwidth}{\kern6pt#1\vskip6pt}\kern6pt\vrule}\hrule}
        \end{center} \vskip-5mm}
\def\lboxit#1{\vbox{\hrule\hbox{\vrule\kern6pt
      \vbox{\kern6pt#1\vskip6pt}\kern6pt\vrule}\hrule}}

\def\thickboxit#1{\vbox{{\hrule height 1mm}\hbox{{\vrule width 1mm}\kern6pt
          \vbox{\kern6pt#1\kern6pt}\kern6pt{\vrule width 1mm}}
               {\hrule height 1mm}}}

\def\fat#1{\hbox{\rlap{$#1$}\kern0.25pt\rlap{$#1$}\kern0.25pt$#1$}}

\newcolumntype{R}{@{\extracolsep{0.5cm}}r@{\extracolsep{0pt}}}%            
\newcolumntype{E}{@{\extracolsep{0.25cm}}c@{\extracolsep{0pt}}}%

\makeatletter
\newcommand{\distas}[1]{\mathbin{\overset{#1}{\kern\z@\sim}}}%
\makeatother

\usepackage{xr-hyper}
\usepackage[colorlinks, linkcolor=blue, anchorcolor=blue, citecolor=blue]{hyperref}

\newtheorem{prop}[theorem]{Proposition}

\usepackage[left=0.9in,right=0.9in,top=0.95in,bottom=0.95in]{geometry}

\makeatletter
\newcommand*{\addFileDependency}[1]{% argument=file name and extension
  \typeout{(#1)}
  \@addtofilelist{#1}
  \IfFileExists{#1}{}{\typeout{No file #1.}}
}
\makeatother

\newcommand{\blind}{0}

\begin{document}

\if1\blind
{
	\title{\bf Hierarchical effect decomposition and regularization for heterogeneous readmission prediction}
	\author{Ziren Jiang$^{1}$, Jared D. Huling$^{1}$\thanks{corresponding author: huling@umn.edu}\\
		$^{1}$Division of Biostatistics, University of Minnesota \\ [8pt]
	}

	\date{}
	\maketitle
} \fi

\if0\blind
{
	\title{\bf Heterogeneous readmission prediction with hierarchical effect decomposition and regularization}
	\author{Ziren Jiang$^{1}$, Lingfeng Huo$^{1}$, Jue Hou$^{1}$, \\
    Mary Vaughan-Sarrazin$^{2,3}$, Maureen A. Smith$^{4}$,
    Jared D. Huling$^{1}$\thanks{corresponding author: huling@umn.edu}\\
		1. Division of Biostatistics and Health Data Science, \\
        University of Minnesota.\\
        2. Center for Access and Delivery Research and Evaluation, \\
        Iowa City Veterans Affairs Medical Center.\\
        3. Department of Internal Medicine, \\
        Carver College of Medicine, University of Iowa.\\
        4. Department of Population Health Sciences, \\
        University of Wisconsin School of Medicine and Public Health.}
	\date{}
	\maketitle
} \fi

%\pagenumbering{gobble}

\begin{abstract}
Accurately predicting hospital readmission risks using electronic health records (EHRs) is critical for effective patient management and healthcare resource allocation. Patient populations in health systems are highly heterogeneous across different primary diagnoses, necessitating tailored yet interpretable prediction models. We propose a hierarchical modeling framework incorporating hierarchical nested re-parameterization and structured regularization methods, which we call hierNest. Specifically, our approach leverages the inherent hierarchical structure present in primary diagnoses and groupings of these diagnoses into major diagnostic categories. Our methodology facilitates information borrowing across related patient subgroups and preserves interpretability at different hierarchical levels. Simulation studies demonstrate superior predictive accuracy of the proposed method, particularly with small subgroup sample sizes and varying degrees of hierarchical effects. We apply our methods to a large EHR dataset comprising Medicare patients. 
\\[3em]    
\end{abstract}%

\def\spacingset#1{\renewcommand{\baselinestretch}%
{#1}\small\normalsize} \spacingset{1.5}

\newpage
\spacingset{1.725} % DON'T change the spacing!
%\spacingset{1.5} % DON'T change the spacing!
\setlength{\abovedisplayskip}{7pt}%
\setlength{\belowdisplayskip}{7pt}%
\setlength{\abovedisplayshortskip}{5pt}%
\setlength{\belowdisplayshortskip}{5pt}%
%\pagenumbering{arabic}
%\setcounter{page}{1}
%%%%   ------------------------------------
%%%%     Input introduction section
%%%%   ------------------------------------
\section{Introduction}

Quantifying the patient health risk in health system settings, such as risk of avoidable hospital readmissions, is critical for identifying patients in need of intervention \citep{evans2016electronic} and is a critical component of health system performance evaluation \citep{mcilvennan2015hospital}. 
Building risk prediction models for avoidable hospital readmissions is critical not only for patient care, guiding timely interventions and management strategies to improve patient-centered outcomes, but also as a metric for evaluating healthcare system performance. 
Electronic Health Records (EHR) data have become an essential resource for monitoring and enhancing the quality and delivery of healthcare \citet{evans2016electronic, goldstein2016opportunities,hou2023risk}. Such data often include extensive patient-level information from diverse sources, reflecting real-world medical practices and patient outcomes. Utilizing EHR data to develop risk prediction models is an increasingly prevalent strategy to inform clinical decision-making, particularly for identifying patients at risk for adverse outcomes such as hospital readmissions. 
Yet, building effective and accurate health system risk models tailored to the needs of patients is a major challenge given the limited data available in a health system, and the high-dimensional, complex nature of EHRs.

Health system populations are large and highly diverse, compounding challenges in building accurate readmissions risk models for patients across a health system. 
Patients with different medical contexts are likely to have vastly different care needs and risks. Such heterogeneity can lead to differential relationships between patient characteristics and their risk of readmission. For example, blood sugar may play a different role in health risk for a patient hospitalized for heart failure versus hyperosmolar hyperglycemic syndrome. 
Or a clinical factor such as elevated blood pressure may influence readmission risk differently for patients hospitalized for cardiac issues compared to those admitted with physical injuries or burns. As such, building risk models that accommodate this kind of heterogeneity and provide information tailored to the specific health context of an individual can better aid and inform health care decision-making. In readmissions modeling, patient heterogeneity is often well-captured by the diagnosis related group (DRG) of the prior (index) hospitalization \citep{elkin2023diagnosis}, as DRGs are a patient classification system designed to classify the nature of a hospitalization into clinically-coherent groups. DRGs can be highly-specific, related DRGs are further grouped into 25 mutually-exclusive Major Diagnostic Categories (MDCs), which largely constitute major organ systems or categorizations of diagnoses, such as trauma or burns or viral infections.
Risk models explicitly designed to accommodate patient heterogeneity can improve predictive performance and provide more interpretable models, as shown in  \citet{huling2018risk}. However, the methods in \citet{huling2018risk} are not applicable for a large number of mutually-exclusive, risk-modifying factors.
Existing methods for building risk models typically use a one-size-fits all approach and either ignore patient heterogeneity and the medical context of a given index admission altogether or use difficult to interpret black box approaches \citep{jamei2017predicting} that make it challenging for clinicians to use the risk models to inform clinical decision-making and further yield only marginal improvements, if any, to predictive performance \citep{li2020good, christodoulou2019systematic}. 

Patient heterogeneity can be handled in an interpretable manner by constructing risk models within different patient subgroups formed by DRGs, which can number in the hundreds. However, handling such patient heterogeneity is a major challenge with high-dimensional EHRs, as sample sizes within subgroups of interest can be prohibitively small. For example, in our motivating study, we have complex information from socio-demographics diagnoses, chronic condition indicators, lab values, medical procedures, and other relevant clinical variables, measured during a baseline period of up to one year prior to hospital discharge, in total amounting to thousands of covariates per patient. 
The use of subgroup-adaptive models with high-dimensional predictors often results in hundreds of thousands of parameters, many of which are estimated with substantially limited sample size, necessitating both computationally-efficient and data-efficient methods.

To develop computationally viable diagnosis adaptive prediction models with high-dimensional EHR covariates, 
we propose a novel hierarchical re-parameterization of the subgroup-specific covariate effects into an overall common effect, an effect specific to each MDC, and finally a DRG-specific effect. 
Pairing this effect decomposition with regularization allows us to shrink the effects of a covariate across related DRGs towards each other, borrowing strength across related subgroups. If a particular covariate exhibits no heterogeneity across subgroups, the DRG-specific and MDC-specific effects can be shrunk to zero so that only the common effect applies. Similarly, if there are differential effects across different MDCs but not DRGs, then the DRG-specific effects can be shrunk towards zero. A basic implementation of this approach is to utilize the lasso penalty, which shrinks each effect towards zero independently. We show that our reparameterization can be achieved by inputting a modified design matrix into existing software routines for the lasso. However, in some circumstances it may help to further encourage shrinkage of all DRG-specific effects within an MDC towards zero simultaneously. And further, it may not make sense to allow for MDC-specific effects if the common effect is not selected and further may not make sense to allow for DRG-specific effects if the respective MDC-specific effect is not selected. Conversely, if a common effect is zero, all MDC and DRG-specific effects should likely be zero too. Similarly, if an MDC-specific effect is zero, so should the DRG effects below it in the hierarchy. This allows adaptively borrowing of strength across related patient subgroups while preserving the ability to differentiate subgroup-specific risk profiles. Importantly, our approach maintains clinical interpretability at all hierarchical levels—ranging from broad, common covariate effects across all DRGs, to more finely-tailored subgroup-specific effects, providing transparent clinical insights about risk processes. We implement this methodology through an efficient computational framework and software implementation, facilitating practical adoption in large-scale clinical and research settings.  
We demonstrate the application of our methodology using a large dataset from an academic health system in a study of readmission risk prediction. When applied to our motivating dataset with over 60,000 observations and over 1 million parameters, our method tuned with 10-fold cross-validation runs within just a few hours.

The remainder of this manuscript proceeds as follows. Section 2 introduces the methodological framework for our hierarchical re-parametrization and introduces a hierarchical penalization strategy to make further use of the hierarchical structure of the data. Section 3 introduces the computational algorithm for our hierarchical parametrization problem. Section 4 presents comprehensive simulation studies to evaluate our method’s performance relative to existing approaches. An illustrative application of our methodology for hospital readmission is presented in Section 5. Finally, Section 6 discusses our findings and implications for the clinical practice.

%%%%   ------------------------------------
%%%%     Input setup section
%%%%   ------------------------------------
\section{New regularization for hierarchical subgroups in diagnosis-adaptive models}\label{sec:repara}

\subsection{Notation and model setup}

Consider the EHR dataset $\{\mathbf{X}_i,D_i,Y_i\}_{i=1}^n$ where $\mathbf{X}_i$ denotes the $p$-dimensional vector of covariates for participant $i=1,...,n$, $Y_i$ denotes the outcome, and $D_i\in \mathcal{D}$ denotes the subgroup, in our application indicated by the DRG, for the $i$-th participant. In our application, $Y_i$ is a binary outcome of readmission or death within 30 days, i.e. $Y_i=1$ indicates the unplanned readmission or death and $Y_i=0$ otherwise. The variable $D_i\in \mathcal{D}=\{1,...,n_{\mathcal{D}}\}$ denotes the diagnosis-related group (DRG) indices for participant $i$ % \hj{explicitly declare that 977 is the total number of DRGs} 
where $\mathcal{D}$ is the complete set of the DRGs with a total of $n_{\mathcal{D}}$ DRG indices. The covariate vector $\mathbf{X}_i$ is potentially high-dimensional, capturing patient demographics, laboratory results, diagnoses, procedures, and other clinical variables collected during a one-year baseline period up to the discharge. Let $n_d=\sum_{i=1}^n I(D_i=d)$ denote the sample size within each DRG subgroup $d$. For a set $\mathcal{D}$, let $|\mathcal{D}|$ denote its cardinality.  

This paper aims to develop a DRG-specific risk prediction model for hospital readmissions. Specifically, we model
\begin{equation}\label{eq: outcome model}
    %\logit(\mathbb{E}[Y_i|\X_i,D_i=d])=\beta_{0d}+\X_i^T\boldsymbol{\beta_d}
    \textnormal{logit}(\mathbb{E}[Y_i|\mathbf{X}_i,D_i=d])=\mathbf{X}_i^T\boldsymbol{\beta}_d,
\end{equation}
where the coefficients $\boldsymbol{\beta}_d=(\beta_{d,1},...,\beta_{d,p})$ is allowed to vary by DRG, hence enabling diagnosis-adapted effects reflecting heterogeneous readmission risks across patients hospitalized for different clinical conditions. 
The model \eqref{eq: outcome model} is flexible in that it allows the relationship between covariates and risk to vary by each subgroup/DRG $d$. However, many DRGs in our application have very small sample sizes, making estimation challenging in even low, let alone moderate dimensional settings. Therefore, in the next subsection, we leverage
the hierarchical structure inherent in the MDC-DRG system to efficiently utilize information across subgroups while accommodating potential heterogeneity in regression coefficients.

\subsection{Hierarchical nested parametrization for MDC-DRG system}

\begin{figure}[ht]
    \centering
    \caption{Illustration of the hierarchical nested structure of the DRGs and MDCs and how our parameter decomposition aligns with this structure.}
     \includegraphics[width=\textwidth]{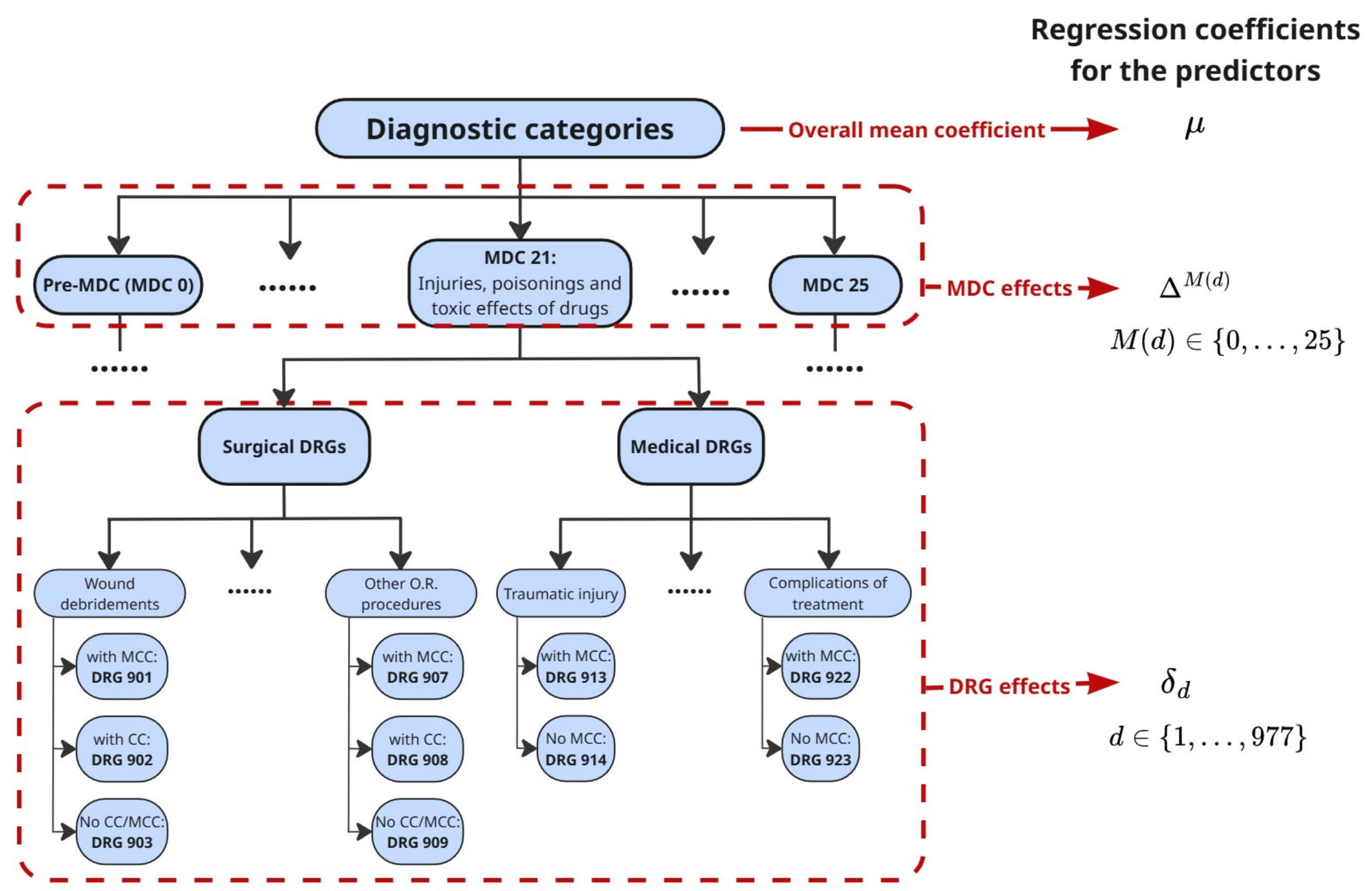}
    \label{fig:MDC_DRG_P1}
\end{figure}

As illustrated in Figure \ref{fig:MDC_DRG_P1}, each diagnosis is initially categorized into one of 26 MDCs. Within each MDC, diagnoses are further subdivided into DRGs, based on: 1) the medical or surgical nature of the condition, 2) specific diagnoses, and 3) the presence of complications or comorbidities. Given this hierarchical classification, it is plausible to assume that DRGs within the same MDC share more similar risk prediction models compared to those from different MDCs. One approach to encouraging such similarity and thus borrowing strength would be to utilize a fused lasso \citep{tibshirani2005sparsity}. However, such penalties such as the fused lasso  are computationally challenging, as they are not separable in the parameters, making their use in our application infeasible. Instead, we re-parameterize the effects in \eqref{eq: outcome model} such that penalization of the terms in the re-parameterization fuses models together in a clinically-sensible manner allowing separable or block separable and thus computationally expedient penalization.
Specifically, we propose re-parameterizing the subgroup-specific covariate effects according to this hierarchical nested structure. We call the resulting modeling and regularization procedure hierNest.

With a slight abuse of notation, we denote $M(d)$ as the specific MDC that the DRG $d$ falls in. Then, we propose the following hierarchical re-parameterization of the regression coefficients for each DRG subgroup $d$:
\begin{equation}\label{eqn:coef_reparam}
\boldsymbol{\beta}_d=\boldsymbol{\mu}+\boldsymbol{\eta^{\textit{M}(d)}}+\boldsymbol{\delta}_d,
\end{equation}
where $\boldsymbol{\mu}$ is the common effect of a covariate across all DRGs, $\boldsymbol{\eta^{\textit{M}(d)}}$ is the additional effect of the covariate that is specific to the MDC group $M(d)$ which shares across all the DRGs that belongs to the same MDC group, and $\boldsymbol{\delta}_d$ is the additional effect specific to DRG $d$.

While \eqref{eqn:coef_reparam} is over-parameterized, we propose estimating the coefficients with penalty terms, which mitigates identifiability issues. By incorporating penalization, the proposed approach enables a structured sharing of information across related DRGs, allowing risk models to be combined flexibly while respecting the hierarchical organization of DRGs within MDCs. For instance, consider two DRG subgroups $d$ and $d'$ within the same MDC, i.e., $M(d)=M(d')$. If the DRG-specific coefficients for the $j$-th covariate, $\delta_{d,j}$ and $\delta_{d',j}$, are penalized to zero, resulting in $\beta_{d,j} = \beta_{d',j}$, this indicates that the $j$-th covariate has the same effect across both DRG subgroups. Such collapsing allows the borrowing of strength, improving estimation.

\subsection{Hierarchical regularization to borrow strength across related subgroups}

We consider estimating model \eqref{eq: outcome model} by a penalized regression objective:
\begin{equation*}
\ell(\boldsymbol{\beta};\mathbf{Y},\mathbf{X}) + P(\boldsymbol{\beta};\boldsymbol{\lambda}),
\end{equation*}
where $\boldsymbol{\beta} = (\boldsymbol{\beta}_d: d\in\mathcal{D})$, $\ell(\boldsymbol{\beta};\mathbf{Y},\mathbf{X})=-\sum_{i=1}^n \bigg[Y_i\mathbf{X}_i^T\boldsymbol{\beta}-\log(1+\exp(\mathbf{X}_i^T\boldsymbol{\beta})\bigg]$ is the negative log-likelihood function, which in our application is chosen to be the negative log of the logistic likelihood, and $P$ is a suitably chosen penalty function with parameter(s) $\boldsymbol{\lambda}$ to be introduced next.
Although a logit link is used here, the modeling framework is readily extendable to alternative distributions and link functions suitable for other outcome types.

Ideally, to borrow strength across related subpopulations, the penalty $P$ should shrink coefficients $\beta_{d,j}$ and $\beta_{d',j}$ closer to each other if subpopulations $d$ and $d'$ share a similar relationship of risk for the $j$th covariate. Further, subpopulations that share the same MDC should have stronger shrinkage as they are more likely to be similar.  
With the hierNest re-parameterization \eqref{eqn:coef_reparam}, we consider penalty terms that allow us to induce the desired coefficient fusion while leveraging more computationally expedient, separable penalties. This will allow our model to data-adaptively traverse between the fully flexible model that has separate coefficients for each DRG and the pooled model where all coefficients are shared in common across all DRGs.

We propose to use different penalization strategies to both 1) collapse related subgroups and 2) incorporate additional prior knowledge about the organization of subgroups into the estimation procedure. Specifically, we propose to use a lasso penalty method (hierNest: Lasso), and an overlapping group lasso penalty method (hierNest: Overlapping Group Lasso) that makes further use of the nested structure of the subpopulations, encouraging additional borrowing of strength across related subgroups.

\subsubsection{hierNest with lasso penalization}

We first propose to use the lasso penalty on the re-parameterized regression coefficients, which adds the penalty $\lambda\sum_{j=1}^p\bigg\{|\mu_j|+\sum_{m\in \mathcal{M}}\bigg(|\eta_j^m|+\sum_{d:M(d)=m}|\delta_{d,j}|\bigg)\bigg\}$ 
to the objective. The incorporation of the lasso penalty enables the collapsing of effects towards the overall effect $\mu_j$, or towards the MDC-specific effects, and allows the DRG-specific effects $\delta_{d,j}$ to be removed or selected individually for all $j=1,...,p$, $M\in \mathcal{M}$, and $d\in M$. 
Because the different terms in the effect decomposition are estimated with distinct subsets of varying sizes, we further adjust the different penalty terms based on their relative sample sizes to ensure relatively equal shrinkage with respect to the amount of information available for the estimation of each term. The proposed lasso penalty with sample size adjustments is:
\begin{equation}\label{eq:lassopenalty}
    P_{l}(\beta)=\lambda\sum_{j=1}^p\bigg\{|\mu_j|+\sum_{M\in \mathcal{M}}\bigg(K^M|\eta_j^M|+\sum_{d\in M}K_d|\delta_{d,j}|\bigg)\bigg\},
\end{equation}
where
\begin{equation*}
        K^M=\bigg(\frac{\sum_{d} n_d}{\sum_{d\in M} n_d}\bigg)^{1/4} \text{ and }
        K_d=\bigg(\frac{\sum_{d} n_d}{ n_d}\bigg)^{1/4}
\end{equation*}
are the penalty factors adjusted for the relative sample size of each subgroup. The adjustment of penalty terms gives higher penalization to the coefficients that correspond to a smaller sample size. Since the overall effect $\mu_j$ corresponds to all participants, the penalty weights penalize the overall effect less compared with the MDC-specific effects or the DRG-specific effects. In a simpler setting, \citet{ollier2017regression} use a similar penalty weight, albeit with the square root of the ratio of the sample sizes. From both our simulations and experience applying our method, we found the fourth root provides more stable penalization in practice.

The lasso penalty \eqref{eq:lassopenalty} allows for borrowing of strength across related subgroups. For example, as the DRG-specific effects $\delta_{d,j}$ within an MDC $M$ are shrunk towards zero, the corresponding overall covariate effects $\beta_{d,j}$ are shrunk towards an effect specific to MDC $M$, i.e. $\beta_{d,j}=\mu_j + \eta_j^M$. Since $\mu_j$ and $\eta_j^M$ are estimated with larger sample sizes, they have higher precision, benefiting the estimation of the effects $\beta_{d,j}$ for smaller subgroups. 

\vspace{-1em}

\subsubsection{hierNest with overlapping group lasso penalization}

Although the sample size adjustment for the lasso penalty terms incorporates information regarding subgroup sample size, it does not put any restrictions on the coefficient structure of effects for a variable overall versus for MDCs and for DRGs. For example, with the lasso penalty \eqref{eq:lassopenalty} for some predictor $j$, the estimated overall effect $\hat{\mu}_j$ can be shrunk to 0 while the DRG-specific effect $\hat{\delta}_{j,d}$ is non-zero for some $d$. In practice, such a violation of effect hierarchy may lead to a large number of false positives. Further, since there are over 900 DRGs in our application, subgroups can be numerous; it is thus important to impose more structure on how covariates can be collapsed to encourage simple models that adhere to the structure of the data. 

We thus propose another penalty that further incorporates the hierarchical structure of the re-parametrized regression coefficients. We first define $\boldsymbol{\theta}_j=\{\mu_j, \{\eta^M_j\}_{M\in \mathcal{M}}, \{\delta_{d,j}\}_{M(d)=M}\}$ as the vector of all coefficients corresponding to the $j$-th predictor, and further define $\boldsymbol{\theta}_j^M=\{\eta^M_j, \{\delta_{d,j}\}_{d\in M}\}$ as the coefficients specific to the MDC subgroup $M$. Then, we propose the following overlapping group lasso penalty:
\begin{equation}\label{eq:OGLassopenalty}
   \begin{split}
        P_{og}(\theta)&=\lambda\sum_{j=1}^p\bigg\{|\boldsymbol{\theta}_j|_2+\sum_{M\in \mathcal{M}}\bigg(|\boldsymbol{\theta}_j^M|_2+|\eta_j^M|_2+\sum_{d\in M}|\delta_{d,j}|_2\bigg)\bigg\}\\
        &=\lambda\sum_{j=1}^p\bigg\{|\boldsymbol{\theta}_j|_2+\sum_{M\in \mathcal{M}}(\alpha_1|\boldsymbol{\theta}_j^M|_2+\alpha_2|\boldsymbol{\theta}_j^M|)\bigg\},
   \end{split}
\end{equation}
%\hj{Definition of $\theta$ is quite involved. Suggest define and explain before (4).}
where $|\cdot|_2$ is the $\ell_2$ norm. Following the results from \citep{jenatton2011structured}, the overlapping group lasso penalty \eqref{eq:OGLassopenalty} can be equivalently expressed in the following way:
\begin{equation*}
    P_{og}(\theta) = \lambda\sum_{G\in\mathcal{G}} |G|_2
\;\text{ where }\;
    \mathcal{G} = \bigcup_{j=1}^p \bigg\{\boldsymbol{\theta}_j, \boldsymbol{\theta}_{j}^{M_1},...,\boldsymbol{\theta}_{j}^{M_{|\mathcal{M}|}}, \eta_{j}^{M_1},...,\eta_{j}^{M_{|\mathcal{M}|}}, \delta_{d_1,j},...,\delta_{d_{|\mathcal{D}|},j}\bigg\}
\end{equation*}
is the overlapping set of penalties. Since the set of allowed zero patterns should be the union-closure of $\mathcal{G}$ \citep{jenatton2011structured}: $\big\{ \bigcup_{G\in\mathcal{G}'} G; \mathcal{G}'\subseteq\mathcal{G}  \big\}$, the proposed overlapping group lasso penalty term enables hierarchical shrinkage for the estimated coefficients. In particular, higher layers of  coefficients (e.g. the common effect of a covariate) can be shrunk to 0 only if all coefficients below it (e.g. the MDC-specific effects) are 0. Thus, the MDC-specific effect $\eta^M_j$ can only be 0 if all the corresponding DRG-specific effects $\delta_{d, j}$ that $M(d)=M$ are 0; and the overall effect $\mu_j$ can only be 0 if all the other subgroup effects for covariate $j$ are 0. 

Unlike the lasso penalty \eqref{eq:lassopenalty}, which enables arbitrary non-zero patterns after the penalization, this overlapping group lasso penalty makes further use of the nested group information and avoids overselection of the DRG-specific effects, which is critical when there are many DRGs and covariates, allowing for simultaneous removal of all DRG- or MDC-specific effects.

%%%%   ------------------------------------
%%%%     Input ED section
%%%%   ------------------------------------

\section{Computation}
In this section, we describe the computational details of the proposed hierNest methods. The hierNest-Lasso problem can be efficiently computed by constructing and supplying a modified design matrix to any existing lasso software, such as \texttt{glmnet}. The hierNest-OGLasso problem, on the other hand, requires specialized computational algorithms. In this section we first describe the modified design matrix for computation for the former and then describe a highly efficient majorization-minimization style algorithm for computation for the latter.

\subsection{Encoding the hierNest decomposition via a modified design matrix}
We first construct the special design matrix with our proposed hierNest re-parametrization to facilitate the use of existing optimization software to fit our penalized, reparameterized model \eqref{eqn:coef_reparam}. Let $\mathbf{X}$ denote the $n\times p$ matrix of the predictors. We define $n_{\mathcal{M}}\coloneqq|\mathcal{M}|$ be the number of MDC groups, and $n_{\mathcal{D}}\coloneqq |\mathcal{D}|$ be the number of DRG subgroups. 
Then, we can define the MDC indicator matrix $\mathbf{H}_{\mathcal{M}}$ as a $n\times n_{\mathcal{M}}$ matrix with the $i$-th row and $j$-th column element being 1 if participant $i$ belongs to the $j$-th MDC group, and 0 if otherwise. Similarily, we can define the DRG-indicator matrix $\mathbf{H}_{\mathcal{D}}$ as a $n\times n_{\mathcal{D}}$ matrix indicating which DRG to which an observation belongs. The hierNest structure matrix $\mathbf{H}$ can be constructed as 
\begin{equation*}
    \mathbf{H} = \bigg[ \mathbf{1}_n; \mathbf{H}_{\mathcal{M}}; \mathbf{H}_{\mathcal{D}} \bigg],
\end{equation*}
where $\mathbf{1}_{n}$ is a $n\times1$ vector of all 1s. Then, the design matrix with the hierarchical nested re-parametrization $\mathbf{X_H}$ is a $n\times p(1+n_{\mathcal{M}}+n_{\mathcal{D}})$ matrix with
\begin{equation}
    \mathbf{X_H} = \bigg( \mathbf{X}^T \odot \mathbf{H}^T \bigg)^T,
\end{equation}
where $\odot$ is the Khatri–Rao product (columnwise Kronecker product) and the superscript $T$ is the transpose operation. It can be shown that fitting a regression model with $\mathbf{X_H}$ as the design matrix and the response vector as the response is equivalent to our parameterization \eqref{eqn:coef_reparam}. The design matrix $\mathbf{X_H}$ is the Kronecker product for each row of $\mathbf{X}$ and $\mathbf{H}$, so we need to transpose those matrices and apply the columnwise Kronecker product (Khatri–Rao product). The structure of the hierNest design matrix $\mathbf{X_H}$ is illustrated in Figure \ref{eq:hierNestmatrix}, where $\mathbf{X}=[\mathbf{X}_1^T;\mathbf{X}_2^T;...;\mathbf{X}_{n_{\mathcal{D}}}^T]^T$ is the partition of the predictors according to the subgroup of DRG. Note that, most of the elements in the hierNest matrix $\mathbf{X_H}$ are 0 due to its structure, allowing its memory storage requirements to be minimal and further accelerating computation that takes advantage of sparse matrices.

\begin{figure}[h!]
\centering
\resizebox{\textwidth}{!}{$
\mathbf{X_H}=
\left(
\begin{array}{c|c|ccc}
\begin{array}{c}
\mathbf{X}_1 \\
\vdots \\
\mathbf{X}_{19}
\end{array}
&
\begin{array}{cccc}
\mathbf{X}_1& \;\;\;\;\;\;\;\; &\;  &\; \\
\vdots&\; & \;  \boldsymbol{0}&\; \\
\mathbf{X}_{19}& \;& \; &\; \\
\end{array}
&
\begin{array}{ccc}
\mathbf{X}_1 & \cdots & \boldsymbol{0} \\
 \vdots & \ddots & \vdots \\
 \boldsymbol{0} & \cdots & \mathbf{X}_{19}
\end{array}
&
  & \boldsymbol{0}
\\
\begin{array}{c}
\mathbf{X}_{20} \\
\vdots \\
\mathbf{X}_{103}
\end{array}
&
\begin{array}{cccc}
\;& \mathbf{X}_{20} &\;\;\; &\; \\
\;& \vdots & \; &\; \\
\;& \mathbf{X}_{103}&\;  &\; \\
\end{array}
&
 &
\begin{array}{ccc}
\mathbf{X}_{20} & \cdots & \boldsymbol{0} \\
\vdots & \ddots & \vdots \\
\boldsymbol{0} & \cdots & \mathbf{X}_{103}
\end{array}
&
 
\\
\begin{array}{c}
\vdots \\
\mathbf{X}_{969} \\
\vdots \\
\mathbf{X}_{977}
\end{array}
&
\begin{array}{cccc}
\;&\;\; &\ddots & \;\\
\;&\; & \;  & \mathbf{X}_{969}\\
\;&\; \boldsymbol{0} & \; & \vdots \\
\;&\; & \; & \mathbf{X}_{977}\\
\end{array}
&
\boldsymbol{0} &   &
\begin{array}{cccc}
\ddots &  &   &   \\
& \mathbf{X}_{969} & \cdots & \boldsymbol{0} \\
& \vdots & \ddots & \vdots \\
& \boldsymbol{0} & \cdots & \mathbf{X}_{977}
\end{array}
\end{array}
\right)
$}
\caption{Block-structured design matrix $\mathbf{X_H}$ with overall, MDC-specific, and DRG-specific effects.}
\label{eq:hierNestmatrix}
\end{figure}

\subsection{hierNest: Lasso}

With the constructed design matrix $\mathbf{X_H}$, the hierNest-Lasso problem can be written as:
\begin{equation}\label{eq:hierNestLassoProblem}
    \min_{\mathbf{\beta_H}}\bigg\{ \ell(\boldsymbol{\mathbf{\beta_H}};\mathbf{Y},\mathbf{X_H}) + \lambda  |K^T\mathbf{\beta_H}|\bigg\},
\end{equation}
where $\ell(\boldsymbol{\mathbf{\beta_H}};\mathbf{Y},\mathbf{X_H})$ is the negative log-likelihood function for logistic regression, $\mathbf{\beta_H}=\bigcup_{j=1}^p\boldsymbol{\theta}_j=\bigcup_{j=1}^p\{\mu_j, \{\eta_j^M\}_{M\in\mathcal{M}},\{\delta_{d,j}\}_{d\in\mathcal{D}}\}$ is the re-parametrized regression coefficients, and $K$ is the sample size adjustment factor defined in \eqref{eq:lassopenalty} correspond to each coefficient. Then \eqref{eq:hierNestLassoProblem} can be efficiently solved by any existing lasso software. We adopt the R package \texttt{glmnet} \citep{glmnet} for solving \eqref{eq:hierNestLassoProblem}. The sample size adjustment factor $K$
can be incorporated with the specification of the ``penalty factor'' input. As most of the elements in $\mathbf{X_H}$ are 0, specifying it as a sparse matrix makes the computation more efficient.

\subsection{hierNest: Overlapping Group Lasso}

\subsubsection{Group-wise majorization-minimization for nested overlapping groups}

The hierNest with overlapping group lasso penalty problem can be written as:
\begin{equation}\label{eq:targetfunction}
    \begin{split}
        &\min_{\mathbf{\beta_H}}\bigg\{ \ell(\boldsymbol{\mathbf{\beta_H}};\mathbf{Y},\mathbf{X_H}) + P_{og}(\mathbf{\beta_H})\bigg\}\\
        & = \min_{\mathbf{\beta_H}}\bigg\{\ell(\boldsymbol{\mathbf{\beta_H}};\mathbf{Y},\mathbf{X_H})+\lambda\sum_{j=1}^p\big\{|\boldsymbol{\theta}_j|_2+\sum_{M\in \mathcal{M}}(\alpha_1|\boldsymbol{\theta}_j^M|_2+\alpha_2|\boldsymbol{\theta}_j^M|)\big\}\bigg\}.
    \end{split}
\end{equation}
To the best of our knowledge, there is no existing software that can directly solve this problem. Therefore, in this subsection, we propose an efficient group-wise majorization-minimization algorithm. 

Denote the partial negative log-likelihood for the $j$-th predictor with coefficients $\boldsymbol{\theta}_j$ as
\begin{equation*}
    \ell_j(\boldsymbol{\theta}_j;\mathbf{Y},\mathbf{X_H})=-
    \sum_{i=1}^n \bigg[Y_i\mathbf{X}_i^{(j)^T}\boldsymbol{\theta}_j-\log(1+\exp((\mathbf{X_H}^T)_i\mathbf{\beta_H})\bigg].
\end{equation*}
where $\mathbf{X}^{(j)}$ is the design matrix that that corresponds to the $j$-th predictor and $\mathbf{X}^{(j)}_i$ is the $i$-th row of $\mathbf{X}^{(j)}$.
Since each predictor $j=1,...,p$ well separates the target function in \eqref{eq:targetfunction}, we iteratively solve the optimization problem for the $j$-th predictor with the following target function:
\begin{equation}\label{eq:focusedlikelihood}
    \begin{split}
        &\ell_j(\boldsymbol{\theta}_j;\mathbf{Y},\mathbf{X_H})+\lambda\bigg(|\boldsymbol{\theta}_j|_2+\sum_{M\in \mathcal{M}}(\alpha_1|\boldsymbol{\theta}_j^M|_2+\alpha_2|\boldsymbol{\theta}_j^M|)\bigg)
    \end{split}
\end{equation}
In the following, we omit the notation of $j$ for simplicity. The second order Taylor expansion for any point $\boldsymbol{\theta}_0$ is
\begin{equation*}
    \ell(\boldsymbol{\theta})=\ell(\boldsymbol{\theta}_0)+(\boldsymbol{\theta}-\boldsymbol{\theta}_0)^T\nabla\ell(\boldsymbol{\theta}_0)+\frac{1}{2}(\boldsymbol{\theta}-\boldsymbol{\theta}_0)^T\mathbf{H_e}(\boldsymbol{\theta}-\boldsymbol{\theta}_0)
\end{equation*}
for all $\boldsymbol{\theta}, \boldsymbol{\theta}_0$ where $\mathbf{H_e}$ is the Hessian matrix for $\ell(\boldsymbol{\theta})$.

For large-scale problems in our paper, to avoid the matrix multiplication, we use the following inequality to construct a majorization-minimization algorithm,
\begin{equation*}
    \ell(\boldsymbol{\theta})\leq\ell(\boldsymbol{\theta}_0)+(\boldsymbol{\theta}-\boldsymbol{\theta}_0)^T\nabla\ell(\boldsymbol{\theta}_0)+\frac{1}{2t}(\boldsymbol{\theta}-\boldsymbol{\theta}_0)^T(\boldsymbol{\theta}-\boldsymbol{\theta}_0),
\end{equation*}
where $t^{-1}$ here is the largest eigenvalue of the Hessian matrix. The majorized version of the optimization target then becomes:
\begin{equation}\label{eq:mmtarget}
    \ell(\boldsymbol{\theta}_0)+(\boldsymbol{\theta}-\boldsymbol{\theta}_0)^T\nabla\ell(\boldsymbol{\theta}_0)+\frac{1}{2t}(\boldsymbol{\theta}-\boldsymbol{\theta}_0)^T(\boldsymbol{\theta}-\boldsymbol{\theta}_0)+\lambda\bigg(|\boldsymbol{\theta}|_2+\sum_{M\in \mathcal{M}}(\alpha_1|\boldsymbol{\theta}^M|_2+\alpha_2|\boldsymbol{\theta}^M|)\bigg).
\end{equation}
In the rest of this section, for the simplicity of presentation, we reorder the parameters in $\boldsymbol{\theta}$ such that $\boldsymbol{\theta}=[\mu; \boldsymbol{\theta}^{M_1};...;\boldsymbol{\theta}^{M_{|\mathcal{M}|}}]$.

The optimal value can be determined by the subgradient equation:
\begin{equation}\label{eq:mmequation}
    \frac{1}{t}\bigg(\boldsymbol{\theta}-(\boldsymbol{\theta}_0-t\nabla\ell(\boldsymbol{\theta}_0))\bigg)+\lambda\boldsymbol{u}+\lambda\alpha_1\boldsymbol{u}^{\mathcal{M}}+\lambda\alpha_2\boldsymbol{v}=0,
\end{equation}
where $\boldsymbol{u}$ is the subgradient of $|\boldsymbol{\theta}|_2$, $\boldsymbol{u}^{\mathcal{M}}$ is the subgradient of $\sum_{M\in \mathcal{M}}(\alpha_1|\boldsymbol{\theta}^M|_2)$, and $\boldsymbol{v}$ is the subgradient of $\sum_{M\in \mathcal{M}}(\alpha_2|\boldsymbol{\theta}^M|)$. More specifically, $\boldsymbol{u}=\frac{\boldsymbol{\theta}}{|\boldsymbol{\theta}|_2}$ if $\boldsymbol{\theta}\neq \boldsymbol{0}$ and $\boldsymbol{u}\in\{\boldsymbol{x}: |\boldsymbol{x}|_2<1\}$ if $\boldsymbol{\theta}=\boldsymbol{0}$, $\boldsymbol{u}^{\mathcal{M}}=[0; \boldsymbol{u}^{M_1};...;\boldsymbol{u}^{M_{|\mathcal{M}|}}]$ where for any $M\in\mathcal{M}$, $\boldsymbol{u}^M=\frac{\boldsymbol{\theta}^M}{|\boldsymbol{\theta}^M|_2}$ if $\boldsymbol{\theta}^M\neq \boldsymbol{0}$ and $\boldsymbol{u}^M\in\{\boldsymbol{x}: |\boldsymbol{x}|_2<1\}$ if $\boldsymbol{\theta}^M=\boldsymbol{0}$, and the first element of $\boldsymbol{v}$ is 0, the $i$-th element of $\boldsymbol{v}$ ($i\neq1$) is $\boldsymbol{v}_i=\textnormal{sign}({\boldsymbol{\theta}}_i)$ if $\boldsymbol{\theta}_i\neq0$ and is anything between $(-1,1)$ if $\boldsymbol{\theta}_i=0$.

Equation \eqref{eq:mmequation} can be solved by the proximal method described in \citet{bach2012optimization}. Define the coordinate-wise soft threshold operator on a vector $\boldsymbol{\gamma}$ and scalar $b$ to be
\begin{equation*}
    (S(\boldsymbol{\gamma},b))_i=\textnormal{sign}(\gamma_i)(|\gamma_i|-b)_+,
\end{equation*}
where $(z)_+=\max\{z,0\}$. Let $\mathbf{S}_1=S(\boldsymbol{\theta}_0-t\nabla\ell(\boldsymbol{\theta}_0),t\lambda)$. We further partition $\mathbf{S}_1$ according to the MDC $\mathcal{M}$ with $\mathbf{S}_1=[\mathbf{S}_{11}; \mathbf{S}_1^{M_1};...; \mathbf{S}_1^{M_{|\mathcal{M}|}}]$ where $\mathbf{S}_{11}$ is the first element of $\mathbf{S}_{1}$, $\mathbf{S}_1^M$ are the elements corresponds to $\boldsymbol{\theta}^{M}$. Then, for any $M\in\mathcal{M}$, define 
\begin{equation*}
    \mathbf{S}_2^M=\bigg(1-\frac{t\lambda\alpha_1}{|\mathbf{S}_1^M|_2}\bigg)_+\mathbf{S}_1^M.
\end{equation*}
The solution for the optimization problem can then be obtained by
\begin{equation}\label{eq:thetahat}
    \boldsymbol{\hat{\theta}}=\bigg(1-\frac{t\lambda\alpha_2}{|\mathbf{S}_2|_2}\bigg)_+\mathbf{S}_2,
\end{equation}
where $\mathbf{S}_2=[\theta_\mu;\mathbf{S}_2^{M_1},...,\mathbf{S}_2^{M_{|\mathcal{M}|}}]$ and $\theta_\mu$ is the first element of $\boldsymbol{\theta}_0-t\nabla\ell(\boldsymbol{\theta}_0)$, corresponding to the overall effect $\mu$.

\subsubsection{Sequential strong rules and the KKT condition}
To accelerate computation, we can develop sequential strong rules to help minimize unnecessary computation. 
Taking the subdifferential of \eqref{eq:focusedlikelihood} and setting equal to zero gives us the following first-order condition:
\begin{equation}
    \nabla\ell(\boldsymbol{\theta})=\lambda\boldsymbol{u}+\sum_{M\in \mathcal{M}}\lambda\alpha_1\boldsymbol{u}^M+\lambda\alpha_2\boldsymbol{v}.
\end{equation}
We have the following proposition for the first-order condition.

\begin{prop}
    Let $\boldsymbol{\Tilde{\theta}_+}=[0; \frac{[\boldsymbol{\theta}^{M_1}]_+}{|\boldsymbol{\theta}^{M_1}|_2}; ...; \frac{[\boldsymbol{\theta}^{M_{|\mathcal{M}|}}]_+}{|\boldsymbol{\theta}^{M_{|\mathcal{M}|}}|_2}]$ where $[\boldsymbol{\theta}^{M_1}]_+$ is the elementwise absolute value of $\boldsymbol{\theta}^{M_1}$.
Then, we have
\begin{equation*}
    |S(\nabla\ell(\boldsymbol{\theta}),\lambda\alpha_1 \boldsymbol{\Tilde{\theta}}+\lambda\alpha_2\boldsymbol{1}_{-1})|_2\leq |\lambda\boldsymbol{u}|_2,
\end{equation*}
where $\boldsymbol{1}_{-1}$ is the vector of all 1s except that the first element is 0. 
\end{prop}

The proof is provided in the Supplementary Material.
With Proposition 1, we have the following inequality
\begin{equation}
    \begin{split}        |S(\nabla\ell(\boldsymbol{\theta}),\lambda\alpha_1\boldsymbol{\Tilde{\theta}}+\lambda\alpha_2\boldsymbol{1}_{-1})|_2&\leq |\lambda\boldsymbol{u}|_2\leq \lambda,
    \end{split}
\end{equation}
since $|\boldsymbol{u}|_2\leq 1$.
For any fixed $\alpha_1$ and $\alpha_2$, define $c(\lambda)=S(\nabla\ell(\boldsymbol{\theta}),\lambda\alpha_1\boldsymbol{\Tilde{\theta}}+\lambda\alpha_2)$. As in \citet{tibshirani2012strong,liang2022sparsegl}, we make the assumption that $c(\lambda)$ is Lipschitz, i.e.
\begin{equation}
    |c(\lambda)-c(\lambda')|_2\leq |\lambda-\lambda'|.
\end{equation}
Therefore, for a sequence of lambda $\{\lambda_1>\lambda_2>...>\lambda_M\}$, we have 
\begin{equation}
    |c(\lambda_m)|_2\leq |c(\lambda_m)-c(\lambda_{m-1})|_2+|c(\lambda_{m-1})|_2\leq (\lambda_{m-1}-\lambda_m)+|c(\lambda_{m-1})|_2.
\end{equation}
If $(\lambda_m-\lambda_{m-1})+|c(\lambda_{m-1})|_2\leq \lambda_m$, which is 
\begin{equation}
    |c(\lambda_{m-1})|_2\leq(2\lambda_m-\lambda_{m-1}),
\end{equation}
we would be able to have $|c(\lambda_m)|_2\leq\lambda_m$ which allows us ignore the entire vector $\boldsymbol{\theta}$ for the $j$-th predictor when updating the algorithm, thus saving computational time. 

\subsubsection{The algorithm}
We now apply the group-wise MM algorithm and the strong rule to form the final algorithm for solving the penalized regression problem with a fixed $\alpha_1$ and $\alpha_2$ and a sequence of lambda $\{\lambda_1>\lambda_2>...>\lambda_R\}$. Similar to \citet{liang2022sparsegl}, we keep track of two sets: the strong set $\mathcal{S}$ and the active set $\mathcal{A}$. The strong set contains all predictor group $j=1,...,p$ such that the strong rule is not violated. The active set contains all groups that have non-zero coefficients at previous values of $\lambda$. The algorithm is described as follows. 

\begin{table}[ht]
\caption{\large\textnormal{Main algorithm for solving the hierNest OGLasso problem.}}
\begin{tabular}{p{0.95\textwidth}}
\toprule
\textbf{Input:} $\mathbf{X}, \mathbf{y}$; MDC set $\mathcal{M}$; DRG set $\mathcal{D}$; tuning parameters $\{\lambda_1, \ldots, \lambda_R\}$ \\
\textbf{Output:} Estimated coefficients $\hat{\boldsymbol{\theta}}$ \\
\midrule
\textbf{Initialize:} $\mathcal{A} \gets \emptyset$, $\mathcal{S} \gets \emptyset$, $\hat{\boldsymbol{\theta}} \gets \boldsymbol{0}$ \\
\textbf{For} $r = 1$ to $R$: \\
\quad Update $\mathcal{S} \gets \mathcal{S} \cup \left\{ j \in \mathcal{S}^c : \left\| S\left(\nabla\ell(\hat{\boldsymbol{\theta}}), \lambda\alpha_1\Tilde{\boldsymbol{\theta}} + \lambda\alpha_2\boldsymbol{1}_{-1} \right) \right\|_2 > \lambda_m \right\}$ \\
\quad \textbf{Repeat:} \\
\qquad For each $j \in \mathcal{A}$, update $\hat{\boldsymbol{\theta}}_j$ using the MM algorithm defined in Equation~\eqref{eq:thetahat} \\
\qquad Update $\mathcal{A} \gets \mathcal{A} \cup \left\{ j \in \mathcal{S} \setminus \mathcal{A} : \left\| S\left(\nabla\ell(\hat{\boldsymbol{\theta}}), \lambda\alpha_1\Tilde{\boldsymbol{\theta}} + \lambda\alpha_2\boldsymbol{1}_{-1} \right) \right\|_2 > \lambda_m \right\}$ \\
\quad \textbf{Until} no new groups are added to $\mathcal{A}$ from $\mathcal{S}$ \\
\quad \textbf{Repeat:} \\
\qquad Update $\mathcal{A} \gets \mathcal{A} \cup \left\{ j \in \mathcal{S}^c \setminus \mathcal{A} : \left\| S\left(\nabla\ell(\hat{\boldsymbol{\theta}}), \lambda\alpha_1\Tilde{\boldsymbol{\theta}} + \lambda\alpha_2\boldsymbol{1}_{-1} \right) \right\|_2 > \lambda_m \right\}$ \\
\quad \textbf{Until} no new groups are added to $\mathcal{A}$ from $\mathcal{S}^c$ \\
\quad Update $\mathcal{S} \gets \mathcal{S} \cup \mathcal{A}$ \\
\bottomrule
\end{tabular}
\end{table}

%%%%   ------------------------------------
%%%%     Input application of ED section
%%%%   ------------------------------------
\section{Simulation}
\subsection{Data generating mechanism}
In the simulation experiment, we generate the predictors $\mathbf{X}$ with one of the following distributions: 1) normal distribution with mean $\mu_d$ and standard deviation $\sigma_d$, 2) Bernoulli distribution with success probability $p_d$, and 3) Poisson distribution with rate $\lambda_d$. In order to reflect the heterogeneity of the distribution of the covariates for different subgroups, we impose a common distribution for the parameters such that each parameter is a random draw from the common distribution, i.e.,
%\begin{equation*}
$\mu_d\sim N(0, {\delta}/{2}), \;\; \sigma_d\sim \text{Gamma}(\alpha=1, \beta={1}/{\delta}), \;\; p_d\sim TN(p=0.5, {\delta}/{10}), \text{ and } \lambda_d= \bigl \lfloor \Tilde{\lambda}+0.5 \bigr \rfloor$,
%\end{equation*}
where $\Tilde{\lambda}\sim N(\lambda=3, {\delta}/{10})$ and we set $\delta=1$ to be the dispersion parameter that controls the heterogeneity level (i.e., a larger $\delta$ value generates a larger difference in the distribution of covariates among subgroups).

The true coefficients $\{\beta_{0d},\beta_{1d},...,\beta_{pd}\}$ are generated with a hierarchical structure, i.e., $\boldsymbol{\beta_d}=\boldsymbol{\mu}+\boldsymbol{\eta^{M(d)}}+\boldsymbol{\delta_d}$, where $M(d)$ is the specific MDC that the DRG $d$ falls. In the simulation experiment, for $j=1,...,p$ and $d\in \mathcal{D}$, we generate each coefficient as
\begin{equation*}
    \mu_j\sim \textnormal{Unif}(-1,1), \;\; \eta^{M(d)}_j=u^{M(d)}_j\times a^{M(d)}_j \eta,  \text{ and } \delta_{d,j}=v_{d,j}\times a^{M(d)}_j \times b_{d,j},
\end{equation*}
where $u^{M(d)}_j \sim \textnormal{Unif}(-1,1)$, $a^{M(d)}_j\sim \textnormal{Bernoulli}(\gamma)$, $v_{d,j}\sim \textnormal{Unif}(-1,1)$, and $b_{d,j}\sim \textnormal{Bernoulli}(\gamma)$. Here, $\gamma$ controls the sparsity of the coefficient with a smaller $\gamma$ value making more coefficients zero. $\eta=1 \textnormal{ or } 0.2$ is the shrinkage parameter that controls the relative magnitude of the DRG-specific effect and MDC-specific effect. Note that, based on the data-generating mechanism, the DRG-specific effect $\delta_{d,j}$ can exist (i.e., non-zero) only if the MDC-specific effect $\eta^{M(d)}_j$ exists which posits the hierarchical parameter structure. The event rate in the simulation experiments is similar to the event rate in our readmission prediction application.

In this simulation experiment, we choose the number of parameters $p = 60$ with a $1:1:1$ ratio of the Gaussian, Bernoulli, and Poisson distributions. The number of MDCs equals $10$ with $4$ DRGs in each MDC (i.e., there are a total of 40 DRGs). We vary the number of observations for each DRG subgroup $n= (30,50, 100, 200)$, the sparsity of the MDC and DRG specific effects $\textnormal{logit}(\gamma)= (-2, -1, 0, 1, 2)$, and the shrinkage parameter $\eta = (1, 0.2)$. Note that, since there are a total of 40 subgroups, even the subgroup sample size $n=30$ yields more than 1000 observations. The binary outcomes $Y_i|\mathbf{X}_i,D_i=d$ are generated from \eqref{eq: outcome model} but with the above coefficients.

\subsection{Comparator methods}
The risk prediction methods under comparison include: (1) hierarchical nested re-parametrization with overlapping group lasso penalty (hierNest-OGLasso), (2) hierarchical nested re-parametrization with lasso penalty (hierNest-Lasso), (3) regularized lasso regression using the R \texttt{glmnet} package (Lasso) \citep{friedman2021package,friedman2010regularization}, (4) regularized lasso regression with grouped covariates using the R \texttt{grpreg} package (GLasso) \citep{breheny2014package,breheny2009penalized}, (5) regularized lasso regression with sparsity in groups using the R \texttt{sparsegl} (SGLasso) package \citep{liang2022sparsegl}, (6) random forest method using the R \texttt{ranger} package \citep{breiman2001random} was also included to assess how well each method performed compared with off-the-shelf machine learning methods. To also incorporate the group information into the comparator methods, we passed the expanded design matrix to the algorithms along with the group information of the MDC and DRG through the argument \texttt{group} whenever the software allowed. For each covariate, we treat each MDC-specific effect as a separate group and the DRG-specific effect as one group. The tuning parameters in each method are selected through the $10$-fold cross-validation algorithm.

Since our outcome variable is a binary variable, we summarize the prediction performance using the area under the receiver operating characteristic curve (AUROC) \citep{hanley1982meaning}. We generate 500 observations for each DRG subgroup according to the same data-generating mechanism and calculate the AUROC using simulated data for each DRG subgroup. The performance metric is the mean AUROC across all DRG subgroups. We are also interested in the worst AUROC across all the subgroups to assess whether methods are able to prevent poor performance for a small number of subgroups.

\begin{figure}[ht]
    \centering
    \caption{Average AUROC across all DRG subgroups for different methods. The sample size $n$ here represents the sample size for each DRG.}
     \includegraphics[width=\textwidth]{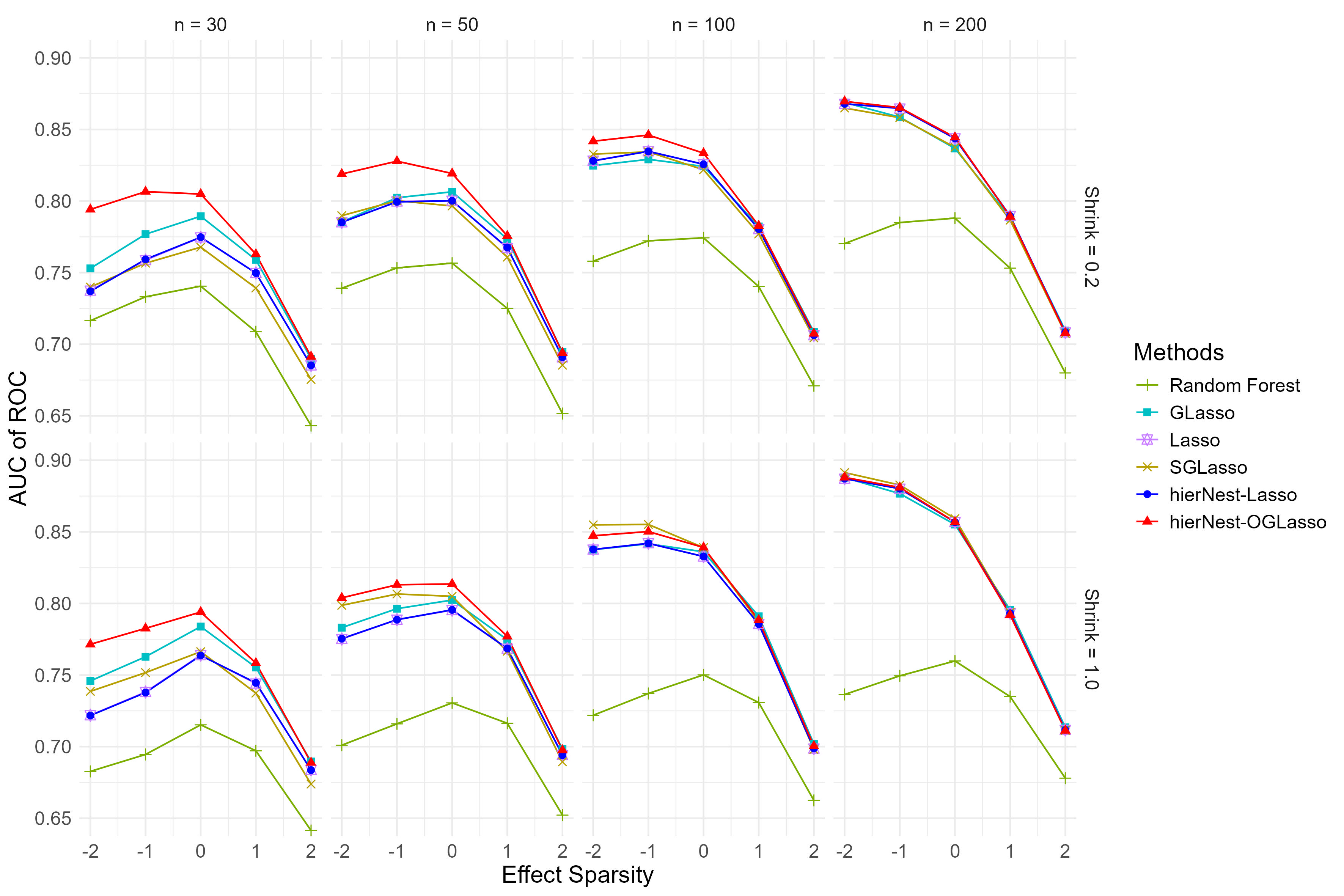}
    \label{fig:AUC_mean}
\end{figure}

\begin{figure}[ht]
    \centering
    \caption{Worst AUROC across all DRG subgroups for different methods. The sample size $n$ here represents the sample size for each DRG.}
     \includegraphics[width=\textwidth]{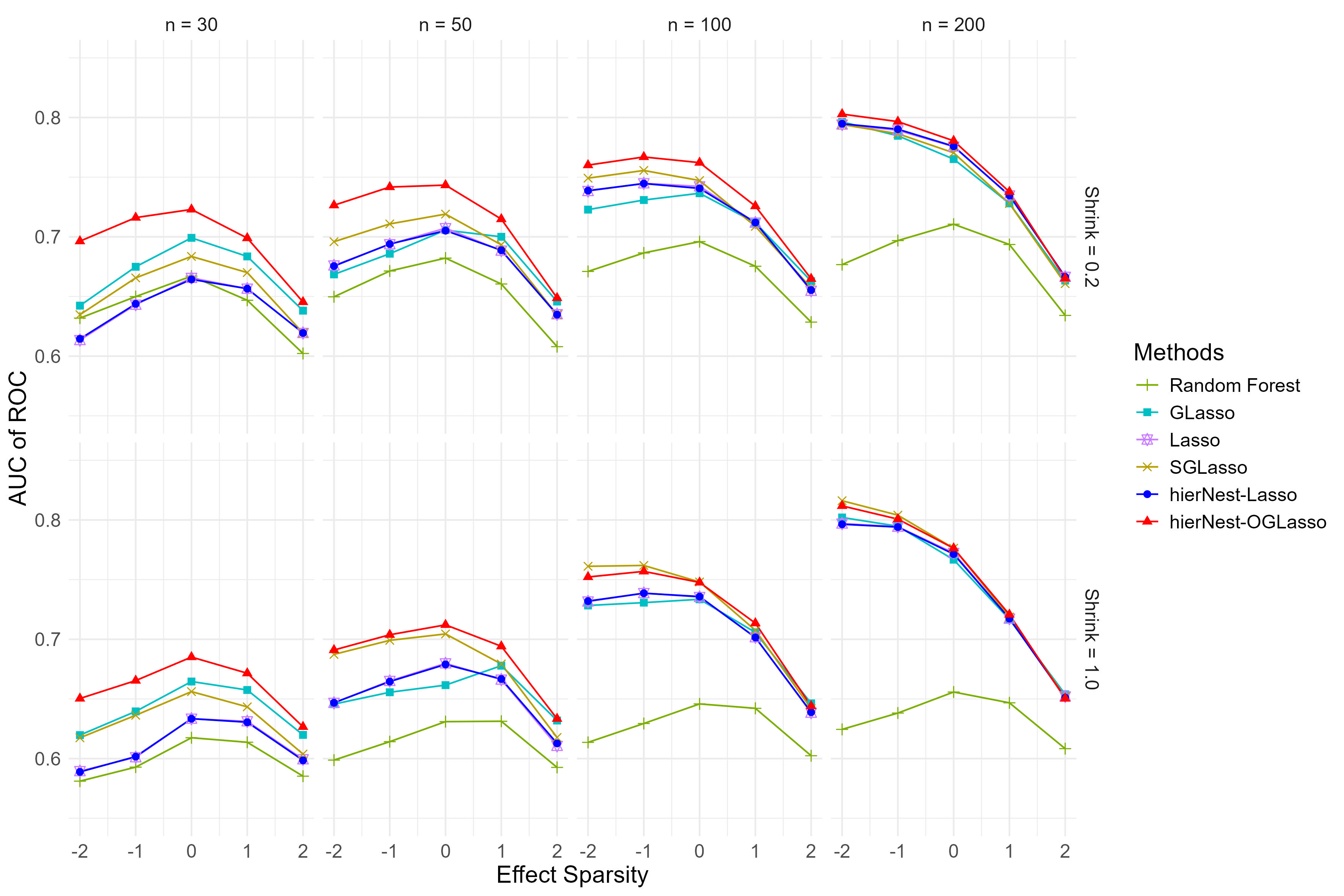}
    \label{fig:AUC_worst}
\end{figure}

\subsection{Simulation results}
Simulation results for the group average AUROC and the worst group AUROC are presented in Figures \ref{fig:AUC_mean} and \ref{fig:AUC_worst} respectively. The results indicate that our hierNest-OGLasso approach consistently outperforms the comparator methods when the shrinkage parameter is set to 0.2 (i.e., the MDC-specific effect is generally smaller than the DRG-specific effect). When the MDC-specific effect is small, ignoring the hierarchical structure of the parameters may incorrectly penalize it to 0. In contrast, our hierNest-OGLasso method always activates the MDC effect as long as its DRG subgroup effects are non-zero. This further validates that our hierNest-OGLasso method benefits from imposing the hierarchical structure of the parameters. 

When the shrinkage parameter is 1, our hierNest-OGLasso method consistently has a better performance when the subgroup sample size is small ($n_d=30$, $n_d=50$ per subgroup). This suggests that our hierarchical re-parametrization, which is designed to leverage information across subgroups, indeed improve the performance when the subgroup sample size is small. As the sample size increases, the performance of the \texttt{sparsegl} algorithm becomes comparable to our hierNest-OGLasso algorithm, with a few cases better. However, the \texttt{sparsegl} algorithm performs significantly worse when the sample size is small or the shrinkage parameter is 0.2. When the sample size is large ($n_d=200$), all the regularized regression methods tend to have a similar performance. 

Our hierNest-Lasso method seems to have similar performance to the general Lasso method, which is suboptimal across all simulation scenarios. We believe this is because we penalize the parameter according to the corresponding sample size. For MDC-specific effect and DRG-specific effect, the penalty factors are high compared to the overall effect $\mu$. This generally degenerates our hierNest-Lasso method to the Lasso method if most of the parameters, except for the $\mu$, are penalized to 0. Therefore, based on the simulation results, we recommend using our hierNest-OGLasso method if the hierarchical structure is plausible.

The simulation results for the worst AUC are displayed in Figure \ref{fig:AUC_worst}, which is consistent with the average AUC results discussed before.

%%%%   ------------------------------------
%%%%     Input ebws section
%%%%   ------------------------------------

\section{Readmission risk prediction in a large health system}\label{sec:casestudy}

The aim of the data application is to construct a risk prediction model in modeling unplanned readmissions of hospitalized patients within 30 days from discharge. We used a retrospective cohort study design with data collected from 2016 - 2020 and outcomes measured following each hospitalization and baseline information collected prior to and including any hospitalization events. The dataset we use includes information for Medicare patients extracted from a Midwestern academic health system's EHR linked with Medicare claims data. Since the dataset is captured on the entire Medicare population and includes all disease categories, we would expect a huge heterogeneity in the prediction model. Therefore, we apply our proposed hierarchical parametrization strategy with different penalty terms to model the unplanned readmissions.

\subsection{Data description}

The original dataset includes a total of 67,804 observations and 7,350 predictor variables. The dataset contains over 736 different DRG groups% \hj{another number of DRG}
. However, we only include  MDCs with more than 30 observations (thus, dropping DRGs with typically fewer than 5 observations). Missing values for the predictor variables are handled by: 1) adding an indicator variable for the missing element, and 2) imputing the missing value with either the median (for continuous variables) or the mode (for categorical variables). All categorical variables are transformed into dummy variables. In order to handle multicollinearity, for the predictor variables that have an empirical correlation larger than 0.95, we only leave the first variable and exclude the others. After the data prescreening, the final dataset we use has a total of 66,885 observations and 3,291 predictors.

In Table \ref{tab:application1}, we plot the number of observations and event rate for each included MDC group. The number of observations has a huge variation among the MDC groups. MDC index 5 (Diseases and disorders of the circulatory system) has the largest sample size in our data, while some other MDC groups have only a few hundreds of observations. However, the event rates are similar among most of the MDC groups.

\begin{table}[ht]
\centering
\caption{Number of observations and 30-day unplanned readmission or death rate by MDC index.}
\small
\renewcommand{\arraystretch}{1.0}
\begin{tabular}{@{\hskip 0pt}p{0.62\textwidth} r r@{\hskip 0pt}} 
\toprule
\textbf{MDC Index} & \textbf{N} & \textbf{Event Rate} \\
\textbf{Total} & \textbf{66,885} & \textbf{0.17} \\
\midrule
MDC 0: Pre-MDC  & 480   & 0.23 \\
MDC 1: Nervous System  & 4,650  & 0.14 \\
MDC 2: Eye   & 29   & 0.10 \\
MDC 3: Ear, Nose, Mouth \& Throat  & 567   & 0.11 \\
MDC 4: Respiratory System & 7,967  & 0.18 \\
MDC 5: Circulatory System & 12,519  & 0.19 \\
MDC 6: Digestive System & 7,396  & 0.18 \\
MDC 7: Hepatobiliary System \& Pancreas & 2,130  & 0.24 \\
MDC 8: Musculoskeletal System \& Connective Tissue & 10,016  & 0.08 \\
MDC 9: Skin, Subcutaneous Tissue \& Breast & 1,636  & 0.17 \\
MDC 10: Endocrine, Nutritional \& Metabolic & 2,426  & 0.20 \\
MDC 11: Kidney \& Urinary Tract & 5,701  & 0.20 \\
MDC 12: Male Reproductive System  & 204   & 0.09 \\
MDC 13: Female Reproductive System & 563   & 0.11 \\
MDC 14: Pregnancy, Childbirth \& the Puerperium  & 37   & 0.19 \\
MDC 16: Blood \& Immunologic & 918   & 0.24 \\
MDC 17: Myeloproliferative \& Poorly Differentiated Neoplasms  & 625   & 0.39 \\
MDC 18: Infectious \& Parasitic Diseases & 6,074  & 0.19 \\
MDC 19: Mental Diseases \& Disorders & 399   & 0.13 \\
MDC 20: Alcohol/Drug Use & 337   & 0.25 \\
MDC 21: Injuries, Poisonings \& Toxic of Drugs & 1,079   & 0.20 \\
MDC 23: Factors Influencing Health Status \& Disorders & 416   & 0.23 \\
MDC 24: Multiple Significant Trauma & 105   & 0.13 \\
MDC 25: Human Immunodeficiency Virus Infections & 611   & 0.23 \\
\bottomrule
\end{tabular}
\label{tab:application1}
\end{table}

\subsection{Model development}

\subsubsection{Methods under comparison and tuning parameter selection}

Similar to the simulation experiments, the prediction performance are evaluated among the following methods: (1) overlapping group lasso with our proposed hierarchical nested re-parametrization (hierNest-OGLasso), (2) lasso with our proposed hierarchical nested re-parameterization (hierNest-Lasso), (3) lasso with the conventional design matrix (lasso), (4) random forest model with the conventional design matrix. We further conduct (5) lasso separately within each MDC alone, which does not share information across MDCs, to serve as the benchmark (lasso-MDCwise); we did not fit an analogous model in a DRG-specific manner as the sample sizes per DRG were often far too small to permit such a modeling strategy. 
The Lasso and random forest methods use the design matrix $\mathbf{X}$, which has 3,291 predictors. The hierNest-OGLasso and hierNest-Lasso approaches use the design matrix $\mathbf{X_H}$ which has a total number of 1,846,812 model parameters. All covariates are standardized prior to fitting the hierNest models.

We use 10-fold cross-validation to select the tuning parameters for each of the risk prediction methods used. Specifically, for the lasso and the hierNest-Lasso methods, we select the penalty parameter $\lambda$ using the built-in cross-validation function \texttt{cv.glmnet()} with 10 folds using area under the receiver operating characteristic curve (AUROC) as the performance measure. For the hierNest-OGLasso method, it has three tuning parameters ($\lambda, \alpha_1,$ and $\alpha_2$) controlling the relative penalization magnitude for the overall effect, MDC-specific effect, and DRG-specific effect. We search over a grid of $\alpha_1,$ and $\alpha_2$; and for each pair of $(\alpha_1, \alpha_2)$, we generate the entire $\lambda$ sequence and evaluate the out-of-fold prediction performance in terms of AUROC. The pair of tuning parameters $(\lambda, \alpha_1,\alpha_2)$ that has the largest mean out-of-fold AUROC is selected as the final tuning parameters.

\begin{figure}[ht]
    \centering
    \caption{Prediction performance within the different validation sets of external hospital observations (external validation) or the Year of 2020 observations (internal observations). We compare the MDC-specific AUPRC(top) and the AUROC(bot) for the compared methods. The black bar represents the median MDC-specific AUCs, and the red bar represents the mean MDC-specific AUCs. Each scatter point represents a single MDC-specific AUC for each MDC subgroup.}
     \includegraphics[width=\textwidth]{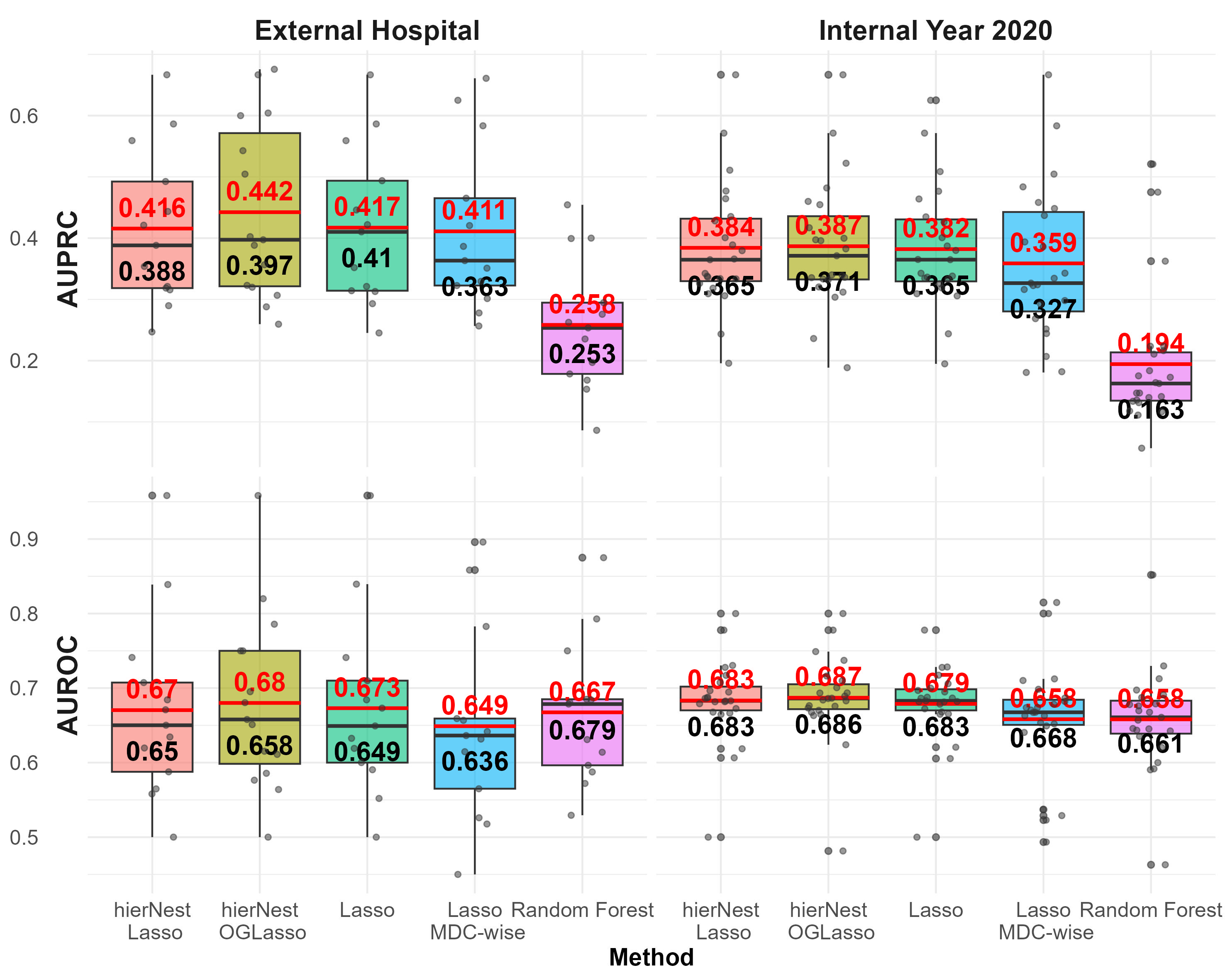}
    \label{fig:res_s}
\end{figure}

\subsubsection{Evaluating prediction performance}

In order to evaluate the prediction performance of the risk models, we separate the entire dataset into the training set (on which the model is fitted and tuning parameters are selected) and the testing set (on which we evaluate the fitted model). We consider the following subset of our data for the testing set to validate the developed risk models: 1) observations from one independent hospital (total of 1,378 observations in the validation set and 65,507 observations in the training set), which has been wholly owned by UW Health for the last 6 years; 2) observations in the year 2020 (total of 13,016 observations in the validation set and 53,869 in the training set).

In addition to the AUROC, we also summarize the prediction performance using the area under the precision-recall curve (AUPRC) \citep{davis2006relationship}. The AUROC measures the model’s ability to discriminate between classes across all possible classification thresholds, providing an aggregate assessment of sensitivity and specificity. The AUPRC, on the other hand, focuses on the model’s performance in terms of precision and recall, which is particularly informative in settings with imbalanced binary outcomes. Because we aim not just to ensure good overall performance, but good performance across as many subpopulations as possible, we evaluate the AUROC and AUPRC for each MDC and consider both the mean and the worst-case AUROC/AUPRC across the MDCs, as it is possible for a model to predict well overall, but have very poor performance for some small subpopulations.

\subsection{Description of final model and results}

\subsubsection{Prediction performance}

In Figure \ref{fig:res_s}, we present the results for the MDC-specific AUROC and AUPRC for the compared methods when using the validation sets of the external Hospital observations and the year 2020 observations. In the boxplots, we present the mean MDC-specific AUCs as red lines and the median as black lines. Each MDC-specific AUCs are plotted as a scatter point along with the box plots. From the figure, our proposed hierNest-OGLasso method has the highest means of both MDC-specific AUROC and AUPRC. Our hierNest-OGLasso method also has a significantly higher 75 percentile of the MDC-specific AUROC and AUPRC with the validation set of external hospital observations. The random forest method consistently exhibits the worst performance across all scenarios, with particularly inferior results in terms of AUPRC. Aside from random forest, the benchmark lasso approach applied separately within each MDC also underperforms relative to the other methods. This outcome is expected, as the method does not leverage information shared across different subgroups. Compared with evaluation using the external hospital validation set, the methods assessed on the 2020 validation data exhibit more similar AUC values, although the hierNest-OGLasso method still demonstrates slightly better performance. This is probably due to the larger sample size for the validation set.
Based on these results, we adopt the hierNest-OGLasso method for the final risk prediction model.

\subsubsection{Final model description}

We describe the final risk prediction model with our proposed hierNest-OGLasso method. We use the model with the validation set of external Hospital observations as the final model since it is developed with a larger sample size. The total number of parameters $\bigcup_{j=1}^{3291} \{\mu_j, \{\eta_j^M\}_{M=1}^{24},\{\delta_{d,j}\}_{d=1}^{536}\}$ in the design matrix $\mathbf{X_H}$ is $1,846,812 = (3291+1)\times(1+24+536)$. The final model has 13,116 coefficients that are estimated with non-zero effects. Among them, only 168 coefficients are the overall effects, 257 coefficients are the MDC-specific effects, and the rest are the DRG-specific effects. In other words, although there are over 13 thousand non-zero coefficients for our over-parametrized model, only 168 of 3,291 predictor variables are eventually selected by our model. Further, 53 out of the 168 selected predictors only have non-zero overall effect (i.e., they have the same effect across all the MDC/DRG subgroups). From the MDC perspective, only 10 out of 24 MDC index (MDC 0: pre-MDC; MDC 1: Nervous System; MDC 4: Respiratory System; MDC 5: Circulatory System; MDC 6: Digestive System; MDC 7: Hepatobiliary System \& Pancreas; MDC 8: Musculoskeletal System \& Connective Tissue; MDC 11: Kidney \& Urinary Tract; MDC 17: Myeloproliferative \& Poorly Differentiated Neoplasms; MDC 21: Injuries, Poisonings \& Toxic Effects of Drugs) have their specific non-zero MDC/DRG coefficients.
That is, our results suggest that diagnoses from all other MDC indexes have the same risk model. In the supplementary material, we listed all the selected predictors along with the number of non-zero coefficients for each MDC.

%%%%   ------------------------------------
%%%%     Input augmented ebws section
%%%%   ------------------------------------
\section{Discussion}\label{sec:discussion}

Hospital readmissions are costly, reflective of poor health outcomes in a health system, yet are often preventable through intervention. Thus, accurate prediction of readmissions is essential for health systems to be better able to target care to patients in need. However, accurately predicting readmissions is highly challenging given the complex nature of health risks across the large, diverse populations served in health systems. In this work we introduced a risk prediction framework that allows flexibility in modeling diverse subpopulations defined by their primary diagnoses but further allows for collapsing the risk model across related subpopulations through a fusion penalty type framework. In our application the subpopulations are defined by DRGs, which are further grouped into MDCs, which constitute related body systems or types of diagnoses. In order to allow fast computation with existing penalized regression software, we use an alternative parameterization of the subpopulation-specific coefficients to allow for fusion of the model across related subpopulations with only a lasso penalty. Our parameterization leverages each covariate's subgroup-specific effect into an overall effect, an effect for MDCs and then an effect specific to the DRG. We further introduce an overlapping group lasso penalty that allows for such fusion while encouraging model hierarchy, only allowing MDC effects when an overall effect is present and only allowing DRG effects when the corresponding MDC effect is present. We found that empirically, incorporation of this structure led to improvements in model performance and interpretability.

Although our motivating application relies on the pre-defined MDC–DRG hierarchy, the proposed approach is not specific to this particular hierarchical structure. Decomposing heterogeneous covariate effects into shared and group-specific components has also been interesting in genetics. For example, \citet{song2024partitioning} considered partitioning genetic effects into cross-tissue (shared) and tissue-specific components, and then aggregates evidence across tissues to improve power for identifying gene–trait associations while still preserving interpretable tissue-level heterogeneity. Our framework is well-suited to settings of this kind, where it is desirable to borrow strength across related groups while allowing for meaningful group-specific deviations.

A natural extension of our framework is to introduce multi-level shrinkage, combining a global regularization parameter with an additional MDC-specific shrinkage parameter to allow the degree of sparsity to vary across MDC groups. This would enable the model to adapt to differences in baseline risk and signal strength. Several limitations should be noted. First, our methods depend on the assumption that the hierarchical structure accurately reflects clinically meaningful relationships, which may not always hold. Future work might explore alternative hierarchical structures or adaptive methods to further optimize model performance under unknown hierarchy or group structure. Another major issue is the limitation of what information is available in the EHR. It is well-known that factors such as social support systems and access to transportation for appointments have a major impact on readmission, yet they are unavailable for use in model-building.

\section{Acknowledgment}
This work was supported through a Patient-Centered Outcomes Research Institute (PCORI) Program Award (ME-2022C1-26326).

\clearpage

%%%%   ------------------------------------
%%%%     Input simulation section
%%%%   ------------------------------------
%\input{section/section_07.tex}

%%%%   ------------------------------------
%%%%     Input case study section
%%%%   ------------------------------------
%\input{section/section_08.tex}

%%%%   ------------------------------------
%%%%     Input discussion section
%%%%   ------------------------------------
%\input{section/section_09.tex}

%\clearpage
%%%%   ------------------------------------
%%%%     Input Appendix
%%%%   ------------------------------------ 
\spacingset{1.5}

\allowdisplaybreaks

\newenvironment{definition}[1][Definition]{\begin{trivlist}
		\item[\hskip \labelsep {\bfseries #1}]}{\end{trivlist}}
\newenvironment{example}[1][Example]{\begin{trivlist}
		\item[\hskip \labelsep {\bfseries #1}]}{\end{trivlist}}
\newenvironment{rmq}[1][Remark]{\begin{trivlist}
		\item[\hskip \labelsep {\bfseries #1}]}{\end{trivlist}}

%\appendix
%
%\setcounter{secnumdepth}{0}
%

%\bigskip
%\begin{center}
%{\large\bf SUPPLEMENTARY MATERIAL}
%\end{center}
%\input{section/section_supp.tex}

%\begin{description}

%\end{description}

\def\spacingset#1{\renewcommand{\baselinestretch}%
{#1}\small\normalsize} \spacingset{0.5}

\bibliographystyle{Chicago}
\bibliography{Bibliography}

@article{song2024partitioning,
  title={Partitioning and aggregating cross-tissue and tissue-specific genetic effects to identify gene-trait associations},
  author={Song, Shuang and Wang, Lijun and Hou, Lin and Liu, Jun S},
  journal={Nature Communications},
  volume={15},
  number={1},
  pages={5769},
  year={2024},
  publisher={Nature Publishing Group UK London}
}

@article{tibshirani2012strong,
  title={Strong rules for discarding predictors in lasso-type problems},
  author={Tibshirani, Robert and Bien, Jacob and Friedman, Jerome and Hastie, Trevor and Simon, Noah and Taylor, Jonathan and Tibshirani, Ryan J},
  journal={Journal of the Royal Statistical Society Series B: Statistical Methodology},
  volume={74},
  number={2},
  pages={245--266},
  year={2012},
  publisher={Oxford University Press}
}

@article{bach2012optimization,
  title={Optimization with sparsity-inducing penalties},
  author={Bach, Francis and Jenatton, Rodolphe and Mairal, Julien and Obozinski, Guillaume and others},
  journal={Foundations and Trends{\textregistered} in Machine Learning},
  volume={4},
  number={1},
  pages={1--106},
  year={2012},
  publisher={Now Publishers, Inc.}
}

@article{glmnet,
    title = {Regularization Paths for Generalized Linear Models via
      Coordinate Descent},
    author = {Jerome Friedman and Robert Tibshirani and Trevor Hastie},
    journal = {Journal of Statistical Software},
    year = {2010},
    volume = {33},
    number = {1},
    pages = {1--22},
    doi = {10.18637/jss.v033.i01},
  }

@article{jenatton2011structured,
  title={Structured variable selection with sparsity-inducing norms},
  author={Jenatton, Rodolphe and Audibert, Jean-Yves and Bach, Francis},
  journal={The Journal of Machine Learning Research},
  volume={12},
  pages={2777--2824},
  year={2011},
  publisher={JMLR. org}
}

@inproceedings{davis2006relationship,
  title={The relationship between Precision-Recall and ROC curves},
  author={Davis, Jesse and Goadrich, Mark},
  booktitle={Proceedings of the 23rd international conference on Machine learning},
  pages={233--240},
  year={2006}
}

@article{hanley1982meaning,
  title={The meaning and use of the area under a receiver operating characteristic ({ROC}) curve.},
  author={Hanley, James A and McNeil, Barbara J},
  journal={Radiology},
  volume={143},
  number={1},
  pages={29--36},
  year={1982}
}

@article{goldstein2016opportunities,
  title={Opportunities and challenges in developing risk prediction models with electronic health records data: a systematic review},
  author={Goldstein, Benjamin A and Navar, Ann Marie and Pencina, Michael J and Ioannidis, John PA},
  journal={Journal of the American Medical Informatics Association},
  volume={24},
  number={1},
  pages={198},
  year={2016}
}

@article{breiman2001random,
  title={Random forests},
  author={Breiman, Leo},
  journal={Machine learning},
  volume={45},
  pages={5--32},
  year={2001},
  publisher={Springer}
}

@article{mcilvennan2015hospital,
  title={Hospital readmissions reduction program},
  author={McIlvennan, Colleen K and Eapen, Zubin J and Allen, Larry A},
  journal={Circulation},
  volume={131},
  number={20},
  pages={1796--1803},
  year={2015},
  publisher={Lippincott Williams \& Wilkins Hagerstown, MD}
}

@article{hou2023risk,
  title={Risk prediction with imperfect survival outcome information from electronic health records},
  author={Hou, Jue and Chan, Stephanie F and Wang, Xuan and Cai, Tianxi},
  journal={Biometrics},
  volume={79},
  number={1},
  pages={190--202},
  year={2023},
  publisher={Wiley Online Library}
}

@article{evans2016electronic,
  title={Electronic health records: then, now, and in the future},
  author={Evans, R Scott},
  journal={Yearbook of medical informatics},
  volume={25},
  number={S 01},
  pages={S48--S61},
  year={2016},
  publisher={Georg Thieme Verlag KG}
}

@article{tibshirani2005sparsity,
  title={Sparsity and smoothness via the fused lasso},
  author={Tibshirani, Robert and Saunders, Michael and Rosset, Saharon and Zhu, Ji and Knight, Keith},
  journal={Journal of the Royal Statistical Society Series B: Statistical Methodology},
  volume={67},
  number={1},
  pages={91--108},
  year={2005},
  publisher={Oxford University Press}
}

@article{breheny2009penalized,
  title={Penalized methods for bi-level variable selection},
  author={Breheny, Patrick and Huang, Jian},
  journal={Statistics and its interface},
  volume={2},
  number={3},
  pages={369},
  year={2009},
  publisher={NIH Public Access}
}

@article{friedman2010regularization,
  title={Regularization paths for generalized linear models via coordinate descent},
  author={Friedman, Jerome and Hastie, Trevor and Tibshirani, Rob},
  journal={Journal of statistical software},
  volume={33},
  number={1},
  pages={1},
  year={2010},
  publisher={NIH Public Access}
}

@article{friedman2021package,
  title={Package ‘glmnet’},
  author={Friedman, Jerome and Hastie, Trevor and Tibshirani, Rob and Narasimhan, Balasubramanian and Tay, Kenneth and Simon, Noah and Qian, Junyang},
  journal={CRAN R Repositary},
  volume={595},
  year={2021}
}

@article{breheny2014package,
  title={Package ‘grpreg’},
  author={Breheny, Patrick and Zeng, Yaohui and Kurth, Ryan and Breheny, Maintainer Patrick},
  journal={URL https://pbreheny. github. io/grpreg},
  year={2014}
}

@article{liang2022sparsegl,
  title={sparsegl: An r package for estimating sparse group lasso},
  author={Liang, Xiaoxuan and Cohen, Aaron and Heinsfeld, Anibal Sol{\'o}n and Pestilli, Franco and McDonald, Daniel J},
  journal={arXiv preprint arXiv:2208.02942},
  year={2022}
}

@article{jamei2017predicting,
  title={Predicting all-cause risk of 30-day hospital readmission using artificial neural networks},
  author={Jamei, Mehdi and Nisnevich, Aleksandr and Wetchler, Everett and Sudat, Sylvia and Liu, Eric},
  journal={PloS one},
  volume={12},
  number={7},
  pages={e0181173},
  year={2017},
  publisher={Public Library of Science San Francisco, CA USA}
}

@incollection{elkin2023diagnosis,
  title={Diagnosis-related group ({DRG})},
  author={Elkin, Peter L and Brown, Steven H},
  booktitle={Terminology, ontology and their implementations},
  pages={379--393},
  year={2023},
  publisher={Springer}
}

@article{li2020good,
  title={How good is machine learning in predicting all-cause 30-day hospital readmission? Evidence from administrative data},
  author={Li, Qing and Yao, Xueqin and {\'E}chevin, Damien},
  journal={Value in Health},
  volume={23},
  number={10},
  pages={1307--1315},
  year={2020},
  publisher={Elsevier}
}

@article{huling2018risk,
  title={Risk prediction for heterogeneous populations with application to hospital admission prediction},
  author={Huling, Jared D and Yu, Menggang and Liang, Muxuan and Smith, Maureen},
  journal={Biometrics},
  volume={74},
  number={2},
  pages={557--565},
  year={2018},
  publisher={Wiley Online Library}
}

@article{christodoulou2019systematic,
  title={A systematic review shows no performance benefit of machine learning over logistic regression for clinical prediction models},
  author={Christodoulou, Evangelia and Ma, Jie and Collins, Gary S and Steyerberg, Ewout W and Verbakel, Jan Y and Van Calster, Ben},
  journal={Journal of Clinical Epidemiology},
  volume={110},
  pages={12--22},
  year={2019},
  publisher={Elsevier}
}

@article{ollier2017regression,
  title={Regression modelling on stratified data with the lasso},
  author={Ollier, Edouard and Viallon, Vivian},
  journal={Biometrika},
  volume={104},
  number={1},
  pages={83--96},
  year={2017},
  publisher={Oxford University Press}
}

\end{document}